# Micro-structural characteristics and indentation behavior of Ce-Al-Ga alloys

*Dharmendra Singh[a], Devinder Singh[b] and R.S.Tiwari[c]*

[a]*Department of Computational Sciences, Central University of Punjab, Bathinda, 151001*
[b]*Amity School of Applied Sciences, Amity University, Lucknow, 226028*
[c]*Nanoscience and Technology Unit, Department of Physics, Banaras Hindu University Varanasi. 221005*

The first report of phase separation in metallic glass has attracted significant attention due to their unique micro-structural variations of amorphous phases at different domain size. In the continuation of this the view is to understanding the genesis of phase separation in Ce based metallic glass. In this presentation, we present extensive investigations with particular reference to low concentration of Ga in Ce-Al-Ga metallic glass. Micro-structural characteristics and indentation behavior of melt spun $Ce_{75}Al_{25-x}Ga_x$ metallic glass has been investigated by X-ray diffraction (XRD) and Transmission electron microscopy (TEM). The small amount concentration of Ga substitution has caused to appearance of second diffuse halos in the X-ray diffraction (XRD) pattern. The observation of "bi-amorphous phases" is thus found in Ce-Al-Ga metallic glass. Indentation characteristics of these metallic glasses have been also investigated. It has been observed that Ga substitution improved the micro hardness property of Ce-Al-Ga alloys. Shear bands around the indentation periphery has also been observed.

*Corresponding author at: Department of Computational Sciences, Central University of Punjab, Bathinda, 151401, India. Tel.: +91–9651190688.
E-mail address: dharmendrasingh4432@gmail.com (Dharmendra Singh)

**1. Introduction:** In recent years, rare earth-based metallic glasses have attracted much research attention due to their unique characteristics. They refer to glass forming ability [1], mechanical [2-8], magnetic properties [9-11]. A large number of novel rare earth based metallic glasses, such as La-, Ce-, Er-, Y-, Sm- based metallic glasses have been synthesized [12]. Out of these, Ce-based metallic glasses show some unique properties due to their unusual behavior linked to 4f electrons [13]. Cerium is the most plentiful rare earth metal on earth. One of the intrinsic characteristic of Ce is its variable electronic structure and valance states [14]. Therefore, the structural and mechanical properties of Ce-based metallic glasses may have characteristics which are different from other known metallic glasses [15]. Recently, effect of cooling rate in mechanical behavior of Ce based metallic glass ribbon has been reported [16-17]. Further studies are necessary to understand the micro-structural features and indentation characteristics in metallic glasses. Keeping these in view, we report the micro-structural and indentation characteristics of $Ce_{75}Al_{25}$ metallic glass composition. Since Ga and Al are isoelectronic with similar atomic radii, the substitution of Al by Ga does not change the e/a ratio of $Ce_{75}Al_{25}$ alloy system. Our previous studies on the nature of Ga systems in various alloy compositions are also relevant in this connection [18-19].

In the present investigation, $Ce_{75}Al_{25-x}Ga_x$ (x=0 and 2) amorphous alloys have been synthesized. The aim is to examine the influence of Ga on the micro-structural and indentation characteristics of $Ce_{75}Al_{25-x}Ga_x$ melt spun glassy ribbons. It will be shown that 2 at. % of Ga for Al changes the amorphous nature drastically. The Ga substitution (for x=2) has led to additional halo peak in the X-ray diffraction (XRD) pattern. The two halos are due to the presence of "bi-amorphous phases". It will be very interesting to understand how a "bi-amorphous phases" can affect the mechanical behavior of these alloys.

**2. Experimental details:** High purity Ce (99.9 %), Al (99.96 %) and Ga (99.99 %) were used for the preparation of alloy ingots of compositions $Ce_{75}Al_{25-x}Ga_x$ (x= 0, 2, 4 and 6 at %) by melting the ingredients in the desired ratios in a silica crucible using a RF induction furnace. The ingots were re-melted several times to improve homogeneity. To convert the ingots into ribbons, they were placed in a silica nozzle tube with a circular orifice of ~ 1 mm diameter. These alloys were then melt spun onto a Cu-wheel rotating at a speed of 40 m/s. The ribbons were prepared by flowing Ar gas continuously. This was done to prevent oxidation of the ribbons after ejection from the nozzle. The length and thickness of the ribbons were ~ 1 to 3 cm and ~ 40 μm respectively. The structural characterization was done by employing X–ray diffractometer (X'Pert Pro PANalytical diffractometer) with $CuK_\alpha$ radiation. The as–synthesized ribbons were thinned using an electrolyte (70% methanol and 30% nitric acid) at 253 K. The thinned samples were then observed under transmission electron microscopy (TEM) using FEI: Technai $20G^2$ electron microscope. Microhardness measurements of all the as-synthesized samples were done with the help of SHIMADZU HMV-2T microhardness tester at different loads. The tests were conducted up to a load till cracks around the indentation impression were observed.

**3. Results and Discussion:**

*3.1 Structural and micro-structural features*

Fig.1 shows the XRD patterns of the $Ce_{75}Al_{25-x}Ga_x$ (0-2 at. %) melt spun ribbons. The XRD pattern shows the broad diffraction maxima, characteristic of amorphous phase without any detectable sharp peaks. For x=0, the broad halo peak is found within the angular range $28^o$ to $35^o$ while for the alloy with x=2, broad halo peak is found within the angular range $39^o$ to $50^o$ in addition to a weak broad peak in the range $28^o$ to $35^o$. The observation of two diffuse halos suggests that there are two types of amorphous phases which are resulting due to phase separation in the alloys $Ce_{75}Al_{25-x}Ga_x$ with x=2. The "bi-amorphous phase" formation seems to be different from the polyamorphic phase transition observed by Yavari et al. in $Ce_{75}Al_{25}$ alloy where a low density amorphous (LDA) phase transforms to high density amorphous (HDA) phase due to application of high pressure [18]. However, it is interesting to note that a small amount of Ga (i.e. 2 at %) drastically reduces the intensity of amorphous halo which is present in the alloy x=0 and produces a rather more strong halo at the higher angle side.

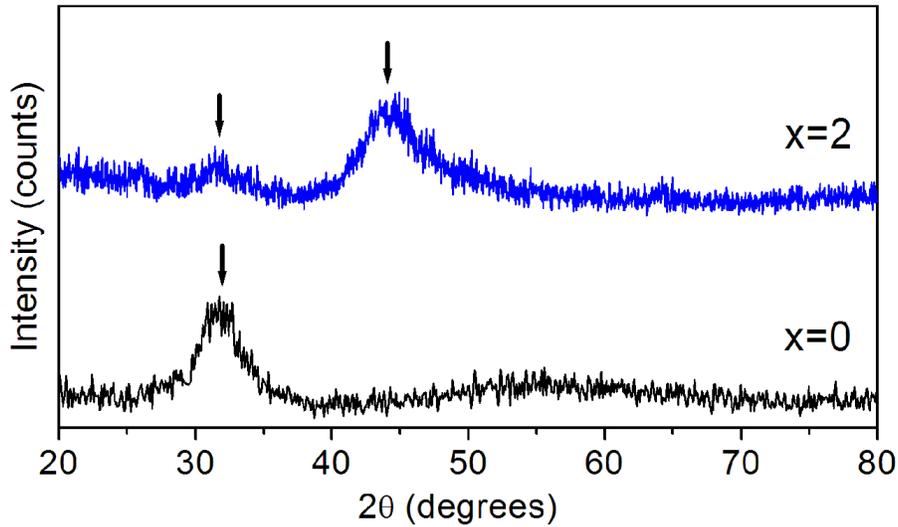

**Fig.1** *XRD patterns of as-synthesized ribbons of $Ce_{75}Al_{25-x}Ga_x$ alloys (x=0 and 4).*

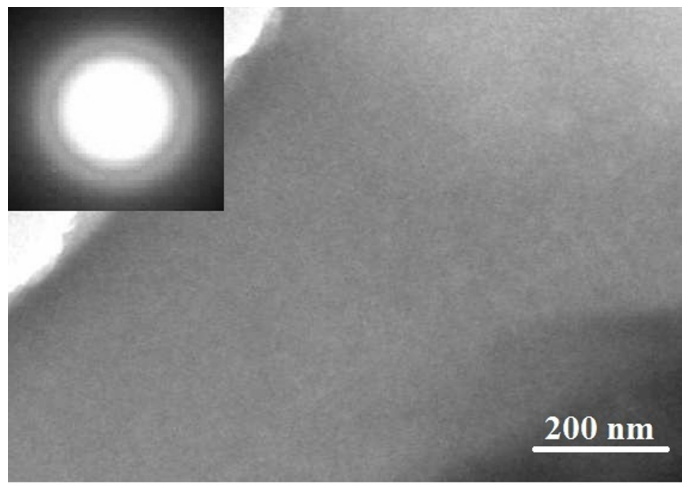

**Fig.2** *A representative Bright field TEM microstructure of $Ce_{75}Al_{25}$ alloy showing the formation of amorphous phase. Inset shows corresponding Selected Area Diffraction Pattern (SADP).*

The formation of amorphous phase in these samples was also confirmed by TEM. Fig.2 and the inset therein shows the TEM micrograph and corresponding selected area diffraction (SAD) pattern displaying diffuse halos for the alloy with x=0. The TEM bright field micrograph for x=0 displays absence of contrast.

*3.2 Indentation characteristics*

The microhardness measurements were carried out at different loads with the help of Vickers hardness tester. The mean hardness reported here is the average of at least five points on each sample. Fig. 3 shows the optical image of the indented samples for the melt spun alloys for $Ce_{75}Al_{25}$ alloy. Optical micrographs revealed that the indentation impressions are regular and crack free at load up to 50 g load for the $Ce_{75}Al_{25}$ alloy.. In addition, Fig. 3 shows the presence of finely spaced wavy patterns at different loads displaying initiation and propagation of shear bands. The deformation seems to occur by the evolution of shear bands [20]. This deformation feature is quite similar to those of bulk metallic glasses [21]. There is no significant variation observed in the formation of shear bands for the melt spun alloy.

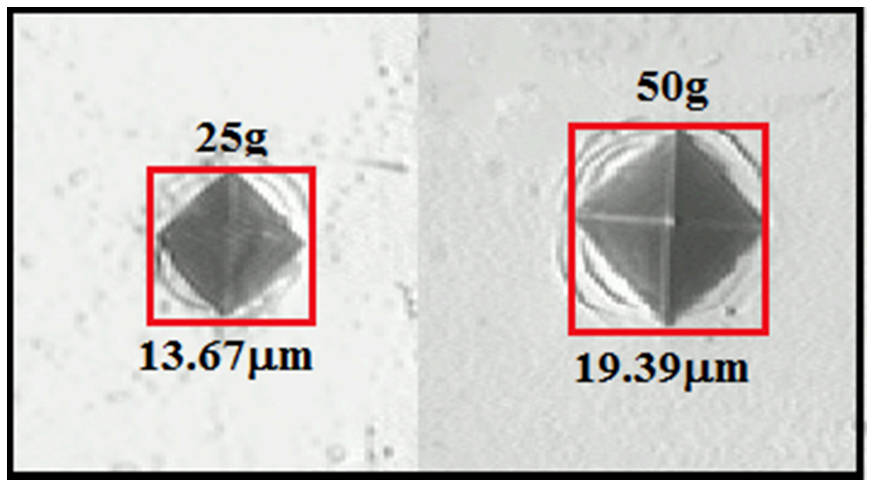

***Fig.3*** *Nature of indentation at different loads for the as-synthesized ribbons of $Ce_{75}Al_{25}$ alloy showing the formation of shear bands around the indentation periphery. Two indentation impressions from various regions of the sample are superimposed.*

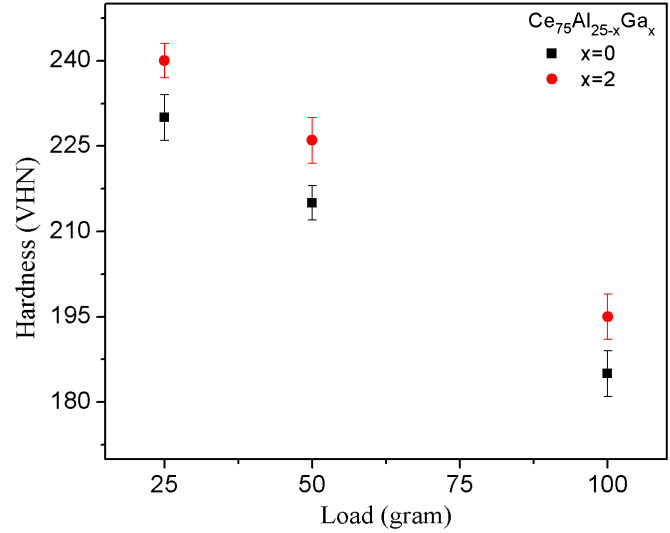

***Fig.4*** *Variation of hardness (VHN) with respect to load (g) of $Ce_{75}Al_{25-x}Ga_x$ alloys (x = 0 and 2).*

The hardness (H) was computed by the formula in GPa units [22];

$$H = 1.854 \times 9.8 \times \frac{P}{d^2} \qquad \ldots(1)$$

where, P is the load (g) and d is the diagonal length in μm. Fig.4 shows hardness versus load characteristic of $Ce_{75}Al_{25-x}Ga_x$ melt spun alloys for x=0 and 2 respectively. The load dependent is seen in this figure. The figure shows that the hardness decreases with increase in the load for all the samples due to indentation size effect [23]. Table 1 presents hardness values at 100g of load for all the melt-spun alloys. We clearly observe increase in hardness with respect to Ga substitution. All other relevant parameters that could be inferred based on hardness versus load curves are given in the table. In the present case, the increase in the hardness of the metallic glasses may be attributed to the variation of free volume with Ga substitution. The Ga substitution may increase the packing of coordination polyhedral and thus would lead to the decrease in the free volume. Such an observation pertaining to the increase in the hardness of metallic glasses with alloying addition has been reported earlier [24].

**Table 1** *Values of hardness (VHN), Meyer's exponent (n), material constant (K) and yield strength ($\sigma_0$) of as-synthesized ribbons of $Ce_{75}Al_{25-x}Ga_x$ alloys.*

| x (in at. %) | VHN (GPa) at 100 gram load (± 0.10) | n | Log K | $\sigma_o$(GPa) at 100 gram load (± 0.10) | VHN/ $\sigma_o$ |
|---|---|---|---|---|---|
| 0 | 1.85 | 1.78 | 1.63 | 1.02 | 1.74 |
| 2 | 1.95 | 1.79 | 1.72 | 1.04 | 1.87 |

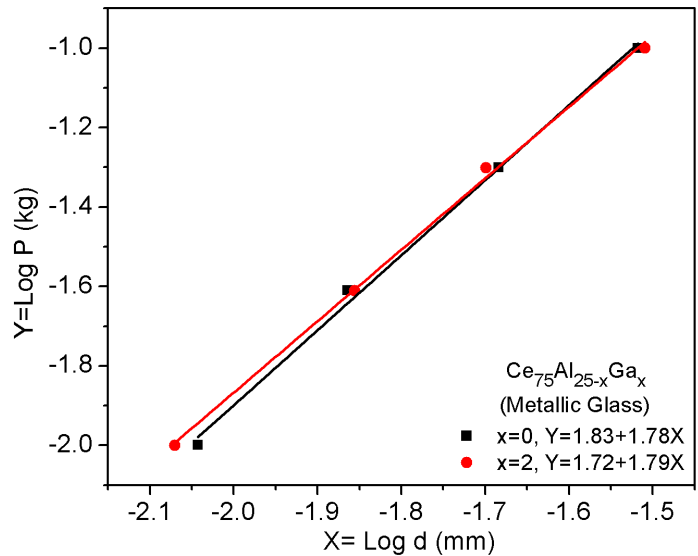

**Fig.5** *Log P versus. Log d plots for the as synthesized $Ce_{75}Al_{25-x}Ga_x$ alloys (x = 0 and 2).*

The hardness values permit us to calculate the 0.2 % offset yield strength. This can be determined by employing the following relationship [25];

$$\sigma_0 = (VHN/3)0.1^{n-2} \qquad \ldots\ldots(2)$$

Where, $\sigma_0$=0.2% offset yield strength and n=Meyer's exponent. The empirical relationship between the applied load P (in Kg) and size of the indentation d (in mm) is called Meyer's law and is given by,

$$P=Kd^n \qquad \ldots\ldots(3)$$

Where n is a material constant related to strain hardening of the material system also known as Meyer exponent. This can be determined from log P versus log d curves. The slope gives n, whereas the intercept K relates to a material constant pertaining to the resistance against penetration by the indenter. Fig. 5 shows Log P versus Log d curves for the melt spun alloys of x=0 and 2. The values of exponent 'n' are given in table 1. The change in the values of n and Log K with increasing value of x has been observed. The values of n are less than 2, as observed for intermetallics [26]. Both n and K increases with Ga concentration and are found to be maximum. The calculation based on equation (2) yield strength increases with Ga addition. The strength of metallic glasses is closely related to the chemical and physical properties of the various elements present in the alloy [27]. Atomic configuration and chemical bonding forces are responsible for origin of strength of the metallic glass.

As discussed earlier, the substitution of Ga in Ce-Al alloy leads to the formation of strong "hump" at the higher angle side in XRD pattern although a weak "hump" is still present in the earlier position. The "hump" position represents short range order (SRO) in the glassy structure. The presence of two "humps" suggests that there are two types of amorphous phases present with different SRO. This may be due to phase separation in the alloys $Ce_{75}Al_{25-x}Ga_x$ (x=2). The hardness behavior of such materials may not be similar to those of monolithic glasses [28].

**4. Conclusions:** The substitution of Ga results in the formation of additional strong diffuse halo in XRD at the higher diffraction angle indicating the formation of two types of amorphous phases in $Ce_{75}Al_{25-x}Ga_x$ alloys. The hardness value of metallic glass (~ 2.28 GPa) has been found maximum for this alloy at 100 g load. The absence of cracks around the indented area up to 50 g of load suggests better fracture toughness of the glassy alloys. The deformation seems to occur by the evolution of shear bands.

**Acknowledgements:**

We would like to thank Prof. O.N. Srivastava and Prof. R.K. Mandal for giving us permission to conduct the experiments in their laboratories.